\documentclass[useAMS]{mn2e}
\usepackage{graphicx}
\usepackage{amssymb}
\usepackage{lscape}
\usepackage{ulem}
\usepackage{txfonts}

\def\kms{km~s$^{-1}$}

\def\si2{Si\,{\sc ii}}
\def\mg2{Mg\,{\sc ii}}
\def\fe2{Fe\,{\sc ii}}
\def\al2{Al\,{\sc ii}}
\def\zn2{Zn\,{\sc ii}}
\def\c2s{C\,{\sc ii}$^{\star}$}

\title[Gas inflows in barred galaxies]
{The impact of gas inflows on star formation rates and metallicities
in barred galaxies.}

\author[Ellison et al.] {Sara L. Ellison$^1$, Preethi Nair$^2$,  David R. 
Patton$^3$,
Jillian M. Scudder$^1$,  J. Trevor Mendel$^1$,
\newauthor Luc Simard$^4$\\
$^1$ Department of Physics and Astronomy, University of Victoria, Victoria, British Columbia, V8P 1A1, Canada.\\
$^2$ INAF-Astronomical Observatory of Bologna,
Via Ranzani 1, 40127 Bologna, Italy.\\
$^3$ Department of Physics \& Astronomy, Trent University, 
1600 West Bank Drive, Peterborough, Ontario, K9J 7B8, Canada.\\
$^4$ National Research Council of Canada,
Herzberg Institute of Astrophysics, 5071 West
Saanich Road, Victoria, British Columbia, V9E 2E7, Canada
}

\begin{document}

\maketitle

\begin{abstract}

The star formation rates (SFRs) and metallicities of a sample of 294
galaxies with visually classified, strong, large-scale bars are
compared to a control sample of unbarred disk galaxies selected from
the Sloan Digital Sky Survey Data Release 4.  The fibre
(inner few kpc) metallicities of barred galaxies are uniformly higher
(at a given mass) than the unbarred sample by $\sim$ 0.06 dex.
However, the fibre SFRs of the visually classified barred galaxies are
higher by about 60\% only in the galaxies with total stellar mass
M$_{\star}$ $>$ 10$^{10}$ M$_{\odot}$.  The metal enhancement at 
M$_{\star}$ $<$ 10$^{10}$ M$_{\odot}$ without an accompanying increase in 
the SFR may be due to a short-lived phase of early bar-triggered star 
formation in the past, compared to on-going star formation rate enhancements
in higher mass barred galaxies.  There is no correlation between bar length or
bar axial ratio with the enhancement of the star formation rate.  
In order to assess the relative importance of star formation triggered 
by bars and galaxy-galaxy interactions, SFRs are also determined
for a sample of close galaxy pairs.  Both mechanisms appear to be 
similarly effective at
triggering central star formation for galaxies with M$_{\star}$ $> 10^{10}$
M$_{\odot}$.  However, due to the much lower fraction of pairs than
bars, bars account for $\sim$ 3.5 times more triggered central star
formation than interactions.

\end{abstract}


\section{Introduction}

Radial gas flows can alter the characteristic properties of galaxies
on relatively short timescales.  Subsequent changes in the star formation
rates (SFRs), gas-phase metallicities, stellar populations and morphologies
contribute to the overall progression of galaxy evolution.  
The mechanisms responsible for gas flows and the changes that they 
ultimately trigger within galaxies are at the heart of understanding 
galaxy transformations.  In particular,
the relative contributions of secular galaxy evolution, whereby galaxies
develop a prominent bulge in the absence of exterior influence, and
galaxy-galaxy interactions is an ongoing-debate (e.g. Kormendy \& Kennicutt
2004; Weinzirl et al. 2009).

In the case of galaxy mergers, the theory and observations are largely
in agreement.  When two gas-rich galaxies interact, the loss of
angular momentum of gas in the outer disks results in radial inflows,
which are expected to lead to triggered star formation as the gas
surface density increases in the inner regions (Barnes \& Hernquist
1996; Di Matteo et al. 2007; Cox et al. 2008).  
These predictions are supported by
observations of star formation rates that are increased by a factor of
a few in close pairs of galaxies (Kennicutt et al.  1987; Barton,
Geller \& Kenyon 2000; Lambas et al. 2003; Alonso et al. 2004;
Nikolic, Cullen \& Alexander 2004; Ellison et al. 2008a), with a clear
tendency for the triggered star formation to be located in the
galactic centres (Barton et al. 2000; Kewley et al. 2006; Ellison et
al.  2010; Patton et al. 2011).  The level of the SFR enhancement in
close pairs is similar out to $z \sim 1$, despite the global increase
of the volume averaged star formation rates at these earlier epochs
(Robaina et al. 2009; Jogee et al. 2009).  Simulations have also
predicted that the triggered star formation should be more effective
in interactions between approximately equal mass galaxies (Cox et al. 2008).  
Again,
this is borne out by observations (Ellison et al. 2008a; Woods et
al. 2006).  Including chemical enrichment into the models leads to
predictions
that, in addition to triggering star formation, the transport of
metal-poor gas from the outer parts of the galaxy results in an
initial dilution of the central gas-phase metallicity (Montuori et
al. 2010; Rupke, Kewley \& Barnes 2010).  Studies of the luminosity-
or mass-metallicity relations in close pairs of galaxies indeed find
that the metallicities are $\sim$ 0.05 dex lower than in galaxies with no
close companion (Kewley et al 2006; Ellison et al. 2008a; Michel-Dansac
et al.  2008), and that abundance gradients are flatter (Kewley et
al. 2010; Rupke, Kewley \& Chien 2010).

In the absence of an interaction, the most effective mechanism for radial
motions of gas (important for bulge growth) is likely to be the
presence of a galaxy bar.  Simulations predict that bars facilitate
not only inflow, but also outflow, so that mixing and triggered star
formation can result in a more complex picture of chemical enrichment
within the galaxy (Friedli, Benz \& Kennicutt 1994).  Studies of the
star formation rates and metallicities of barred galaxies have yielded
complex results.  For
example, whilst many studies have found an increased star formation
rate in barred galaxies (Hummel et al. 1990; Martin 1995; Hawarden,
Huang \& Gu 1996; Huang et al. 1996), bars are apparently neither
required nor guaranteed to yield high star formation rates (Pompea \&
Rieke 1990; Martinet \& Friedli 1997; Chapelon et al. 1999).  

Enhanced SFRs may depend on the morphology of galaxies, being
apparently prevalent in early type barred galaxies, but largely absent
in late types (Huang et al. 1996; Ho et al. 1997; James, Bretherton \&
Knapen 2009).  It has been suggested that the dependence of SFR
enhancement on morphological type may be due to the generally longer,
stronger (more elongated) bars in the former (Elmegreen \& Elmegreen
1985, 1989; Erwin 2005; Menendez-Delmestre et al. 2007).  Simulations
indicate that only strong bars are effective at funnelling gas to the
inner kpc (e.g Regan \& Teuben 2004).  Indeed, Martin (1995) found
that $\sim$ 70\% of starbursts in their sample had a strong bar.
Martinet \& Friedli (1997) also found that only galaxies with strong
and long bars show star formation enhancement, but even strong bars
may be in a pre- or post-starburst phase with `normal' SFRs. Studies
of enhanced molecular gas concentrations in the inner kpc of barred
galaxies (e.g. Sakamoto et al. 1999; Sheth et al. 2005) paint a
similar picture of an early vs.  late-type galaxy separation, plus
considerable variation in SFRs for a given morphological type.  
Indeed, Jogee et al.  (2005) find that the SFRs of barred galaxies
can vary by an order of magnitude for a given central molecular gas
mass.  Sheth et al. (2005) find that early-type bars have higher
molecular gas masses in their centres than late-types, yet find a notable
population of early-types with no central concentration of molecular
gas.  Sheth et al. (2005) suggest that these are galaxies which have
used up their molecular gas in past starbursts, indicating that the
bar lifetime is long compared to the gas inflow/triggered star
formation timescales (see also Sakamoto et al. 1999).  Other indirect evidence
that bars are not promptly destroyed by the build up of central mass
concentrations comes from the relatively high bar fractions out to $z
\sim 1$ (Sheth et al. 2003; Jogee et al. 2004), old stellar
populations (Perez et al.  2009; Sanchez-Blazquez et al. 2011),
multiple extended episodes of star formation in the bar (Cantin et
al. 2010) and the lack of unbarred galaxies whose molecular
concentrations resemble those in bars (Sakamoto et al. 1999).

The chemical enrichment of barred galaxies is similarly complex.
Abundance gradients are often found to be flatter in barred galaxies than their
unbarred counterparts (e.g. Vila-Costas \& Edmunds 1992; Martin \& Roy
1994).   However, a range of abundance gradients exists amongst the
barred population, including some that are considerably steeper than
unbarred spirals (Edmunds \& Roy 1993; Oey \& Kennicutt 1993; Zaritsky
et al 1994; Considere et al. 2000).  Whilst most previous studies have 
focussed on gas-phase abundances, there have also been a handful of 
studies of the stellar populations.  Perez et al. (2009) find that stellar
metallicity gradients within bars can be either positive or negative.
In a detailed study of two early-type barred galaxies, Sanchez-Blazquez
et al. (2011) find stellar abundance gradients that are flatter in the bar than
the disk.  The diversity of gradients may be
due, at least in part, to a dependence on bar length and galaxy type,
whereby long, strong bars and later types tend to exhibit flatter
gradients (Oey \& Kennicutt 1993; Zaritsky et al 1994; Martin \& Roy
1994; Dutil \& Roy 1999).  An added complication is the presence of
breaks in the gradient, often leading to steeper inner slopes and
flatter values outside of the bar (Martin \& Roy 1995; Roy \& Walsh
1997; Considere et al. 2000).  Simulations predict such dependences,
due to the efficiency and quantity of gas transport and whether or not
star formation accompanies these motions (Friedli et al. 1994; Friedli
\& Benz 1995).  

Characteristic (at a given galactocentric radius) and central
abundances in barred galaxies yield perhaps the most puzzling results.
Henry \& Worthey (1999) found no difference at a given M$_B$ between
barred and unbarred metallicities.  Dutil \& Roy (1999) found that
(both early and late type) barred systems were more metal-poor (by
0.5 dex) at a given M$_B$ than unbarred and weakly barred galaxies.
Considere et al. (2000) also found lower central abundances for a given
M$_B$ in their starbursting barred sample.  However, whilst bar
strength may influence the nuclear star formation (Martin 1995) it is
apparently uncorrelated with central metallicity (Chapelon et
al. 1999).  In contrast to the observations, simulations predict that
star formation should lead to \textit{higher} central abundances
(Friedli et al. 1994; Friedli \& Benz 1995).  High \textit{stellar}
metallicities in the bulges of barred galaxies have recently been
reported by Perez \& Sanchez-Blazquez (2011), but similar enhancements
in the gas-phase abundances have not been convincingly demonstrated.

In this paper we aim to use a large sample of visually classified
barred galaxies to quantify differences in the star formation rates
and metallicities relative to an unbarred control set.  In addition to
sample size, the main advantage of our study is internal consistency
in parameter determination and calibration, as well as a control
sample that is fully representive of the barred galaxies' properties.
We are also, for the first time, able to quantify the SFR and
metallicity enhancements on a galaxy-by-galaxy basis and can hence
search for dependences on structural parameters.  Finally, the
enhanced SFRs of barred galaxies are compared to those in close pairs
of galaxies.  The results yield an insight into the relative
efficiency of star formation triggered by bar-driven gas flows and
galaxy-galaxy interactions.  We adopt a cosmology with values of $H_0
= 70 $ \kms Mpc$^{-1}$, $\Omega_M = 0.3$ and $\Omega_{\Lambda} = 0.7$.

\section{Sample}\label{sample_sec}

\begin{figure*}
\centerline{\rotatebox{270}{\resizebox{12cm}{!}
{\includegraphics{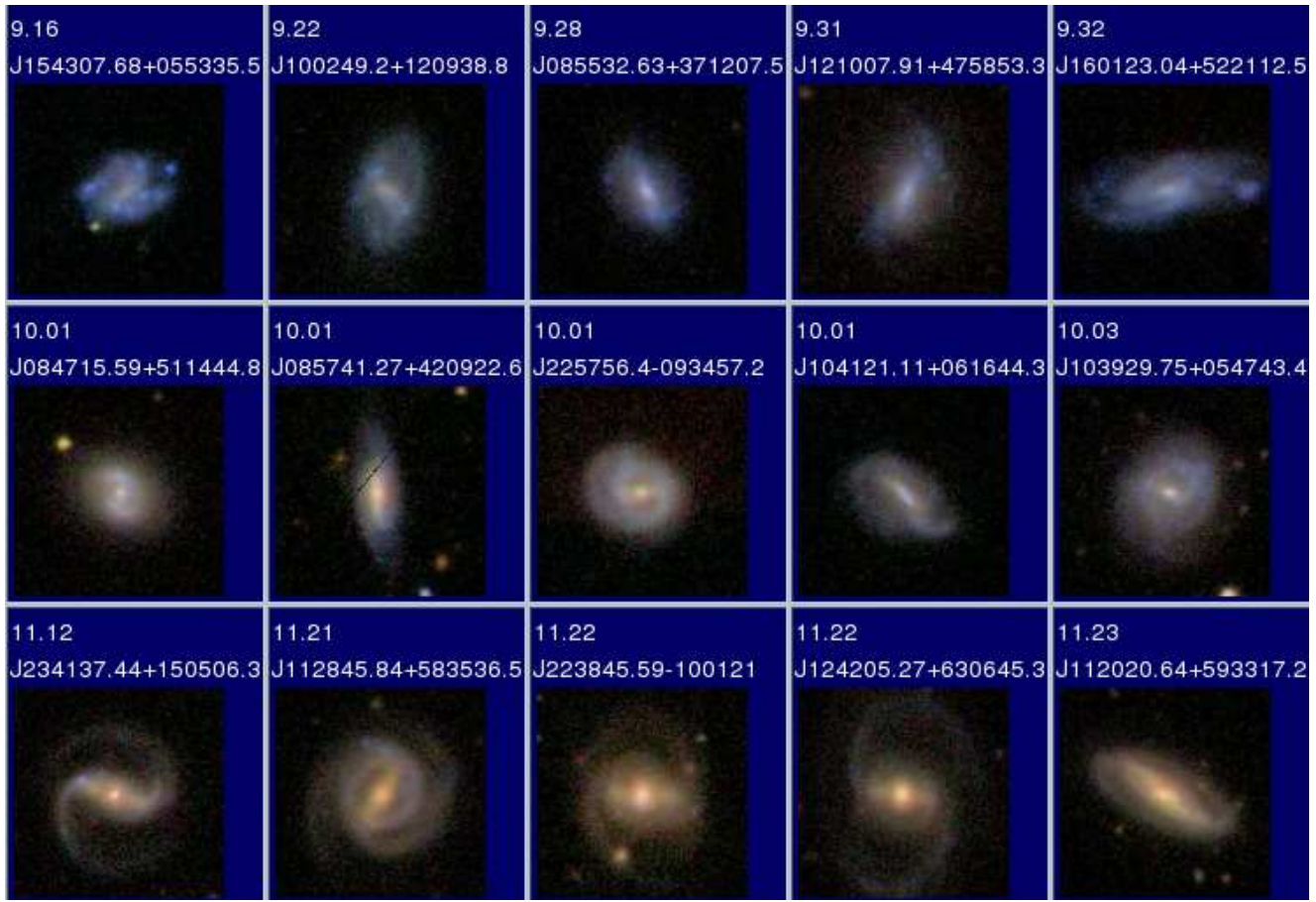}}}}
\caption{\label{bar_images} Top row: The 5 lowest mass barred galaxies in
our sample.  Middle row: 5 galaxies with stellar masses M$_{\star}$ $\sim$
10$^{10}$ M$_{\odot}$.  Bottom row: The 5 highest mass barred galaxies in our sample.
The number at the top of each panel is the stellar mass in units of 
log M$_{\odot}$.}
\end{figure*}

Our sample of barred galaxies is taken from the catalogue of Nair \&
Abraham (2010a).  In brief, Nair \& Abraham (2010a) performed visual
classifications of 14,034 $z<0.1$ galaxies from the spectroscopic
galaxy sample in the data release 4 (DR4) of the Sloan Digital Sky
Survey (SDSS).  In addition to the redshift cut, a magnitude cut was
also imposed such that the extinction corrected SDSS $g$-band modelMag
is brighter than 16.0.  A total of 9917 galaxies are classified as
disks (T Type $\ge -2$), of which 2218 are classified as barred
galaxies (barflag=2,4,8 for strong, intermediate and weak bars
respectively) and 7316 are classified as unbarred (barflag=0). The
remaining 383 galaxies are either classified as `bar unsure', or as
other nuclear structure, such as ansae or peanut morphologies.  Nair
\& Abraham (2010a) note that even their `weak' bars (barflag=8) would
correspond to a strong bar classification in the RC3.  The
classification of strong, intermediate and weak corresponds to the
relative light contribution of the bar, but in all cases the presence
of a bar is considered sure.  See Nair \& Abraham (2010a) for
extensive discussion of their classification process and comparison
with other catalogues.  

Although the bar selection was done with a purely visual
classification, ellipses are fitted post-identification to the bars to
determine their parameters.  The ellipse fitting, which is described
in Section \ref{length_sec}, yields typical bar lengths of 3--10 kpc
with bar axial ratios mostly in the range 0.2--0.6, with a median value of
0.37.  Therefore, although bars can exhibit a range of sizes,
including nuclear and small-scale bars with lengths $\ll 1$ kpc, our
sample is dominated by strong, large-scale bars due to the resolution
of the SDSS images. To remain concise, we will refer to our visually
classified sample as a `barred' sample, with the caveat that the
sample is not complete for small and/or weak bars.

Even for large-scale bars, the effectiveness of visual classifications
diminishes with galaxy inclination, so we further resticted our sample to
have galaxies with axial ratio $b/a \ge 0.4$.  Bars also become
more difficult to identify at higher redshifts.  Nair \& Abraham
(2010a) find a constant bar fraction in their sample out to at least
$z \sim 0.06$.  However, in order to maintain sufficient statistics in
our sample of close pairs (which we use in Section \ref{pairs_sec} to
compare the relative importance of gas inflows from bars and
interactions) we do not impose an \textit{a priori} redshift cut.
Instead, we have checked that none of the results from the barred
sample are affected by a redshift dependence by repeating the analysis
described in Sections \ref{sfr_sec} -- \ref{length_sec} with an upper
redshift cut of 0.06.  Hence we can confirm that there is no effect
from redshift selection in our analysis.

From the list of clearly classified barred and unbarred disk galaxies, we
then select those galaxies with strong emission lines for which
both star formation rates and metallicities can be determined.  Star
formation rates are based on techniques described in Brinchmann et
al. (2004).  Values deduced directly from the spectra (fibre star
formation rates) are corrected to total galactic rates using global
galaxy colours (e.g. Brinchmann et al. 2004; Salim et al. 2007).
Brinchmann et al. (2004) show that their colour-based corrections
essentially remove all aperture effects.
Stellar masses must also be available.  The DR4 catalogue of stellar
masses (Kauffmann et al. 2003) has now been extended to the
DR7\footnote{http://www.mpa-garching.mpg.de/SDSS/DR7/}.  Whereas the
DR4 galaxy masses were derived from the spectra, the DR7 masses 
used here are
derived from fits to the 5-band SDSS photometry.  There is generally
excellent agreement between the two methods.

Metallicities are derived from the SDSS line fluxes available at
http://www.mpa-garching.mpg.de/SDSS/DR7/.  We require high quality
detections of [OII] $\lambda 3727$, H$\beta$ $\lambda 4863$, [OIII]
$\lambda 5007$, H$\alpha$ $\lambda 6563$ and [NII] $\lambda 6584$,
which necessitates a lower redshift cut of 0.02.  Extensive quality
control tests (see Scudder et al. in preparation) were run in order to
ensure reliable line fluxes and our final selection criteria require a
S/N $>$ 5 in all of the lines above.  The measured line fluxes (which
are already corrected for Galactic extinction and underlying
absorption in the stellar continuum) are further corrected for
internal extinction using the ratio of H$\alpha$/H$\beta$ and an SMC
extinction curve (Pei 1992).  Galaxies whose emission line ratios
indicate a significant photoionization contribution from an active
galactic nucleus (AGN) are excluded (Kewley et al. 2001).
Metallicities are calculated based on the re-calibration of the Kewley
\& Dopita (2002) recommended method as presented by Kewley \& Ellison
(2008).

\begin{figure}
\centerline{\rotatebox{0}{\resizebox{8cm}{!}
{\includegraphics{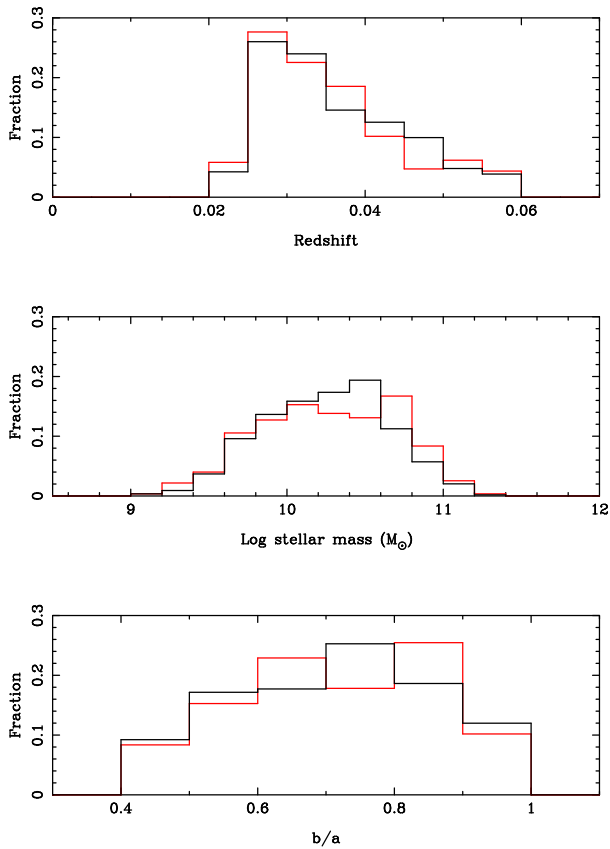}}}}
\caption{\label{mass_z_histo}Fractional distributions of mass,
redshift and galaxy axial ratio (b/a) for the barred (red) and
unbarred (black) samples.  }
\end{figure}

The above criteria result in a final sample of 294 bar galaxies and
806 unbarred galaxies with well-determined metallicities, star
formation rates and stellar masses.  Figure \ref{bar_images} shows the
SDSS postage stamps for a selection of our barred sample.  However,
the barred sample is biased to more face-on galaxies, due to the
difficulty of visually classifying bars in edge-on galaxies.  We
therefore construct an unbiased control sample by finding the unbarred
galaxy which is the best simultaneous match in redshift, stellar mass
and galaxy axial ratio (b/a) without replacement.  The process is
repeated for the second best match and a Kolmogorov-Smirnov test
confirms that the distributions of redshift, mass and b/a of the bars
and control samples are consistent (all probabilities $>$ 25\%).
Figure \ref{mass_z_histo} shows the distributions of stellar mass,
redshift and galaxy axial ratios for the two samples.  Our control
sample size is therefore 588 unbarred galaxies.

\section{Star formation rates}\label{sfr_sec}

There is a strong correlation between SFR and stellar mass (Brinchmann
\& Ellis 2000; Brinchmann et al. 2004; Noeske et al. 2007; Lara-Lopez
et al. 2010a).  In order to search for differences in the SFRs of
barred and unbarred galaxies we fit the SFR-mass relation of unbarred
galaxies with a second order polynomial.  In the first instance, we
consider total SFR and total stellar mass.  For each barred galaxy we
determine a $\Delta$ SFR as the observed total SFR minus the total SFR
predicted for its stellar mass (in log units) based on the fit to the unbarred
population:

\begin{equation}
\Delta \rm{SFR} = \log \rm{SFR} - \log \rm{SFR}_{\rm predict}
\end{equation}

\noindent Positive values of $\Delta$ SFR therefore indicate an enhancement
in the star formation rate of a factor of $10^{\Delta \rm SFR}$.  The
SFR residuals (in bins of stellar mass) of the unbarred control sample
have median values around zero and scatter typically less than 0.05 dex.

\begin{figure}
\centerline{\rotatebox{0}{\resizebox{8cm}{!}
{\includegraphics{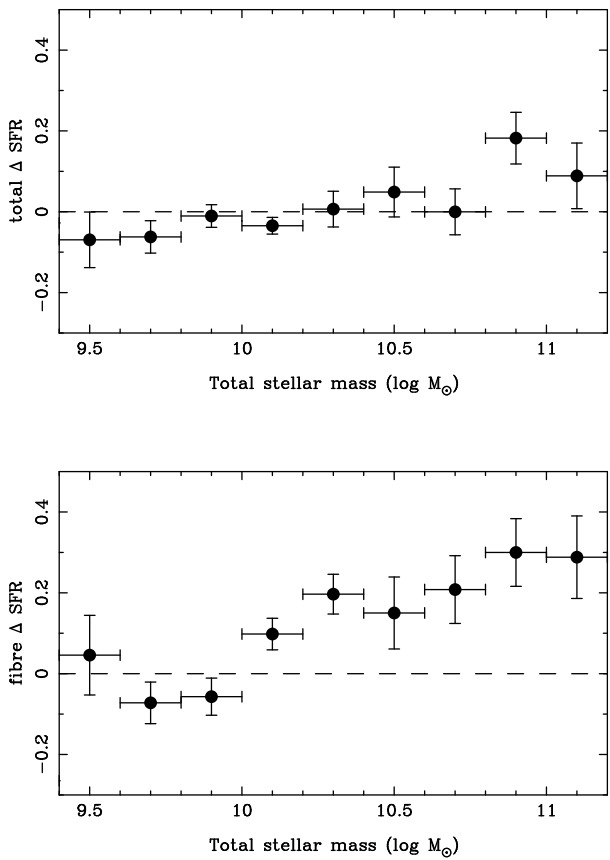}}}}
\caption{\label{delta_sfr} The star formation rate offset of barred
spirals from the SFR-mass relation of unbarred spirals.  The top panel
shows the offset between total SFR in barred and unbarred galaxies at
a given total galactic stellar mass (as a function of total galactic
stellar mass).  The lower panel shows the offset between fibre SFRs in
barred and unbarred galaxies at a given fibre stellar mass (as a
function of total galactic stellar mass). Error bars in this and
all the figures in this paper are the standard error on the mean. }
\end{figure}

In the top panel of Figure \ref{delta_sfr} we plot the $\Delta$ SFR
values as a function of total stellar mass.  Over the range of total
stellar masses in Figure \ref{delta_sfr}, the number of galaxies per
mass bin is $\sim$ 20 (see Figure \ref{mass_z_histo}).  With the
exception of two elevated points at the highest masses (of which only
one has more than 1$\sigma$ significance), there is no evidence for
increased total SFRs in the barred galaxies.  However, simulations of
gas flows in bars (Friedli et al. 1994; Friedli \& Benz 1995; Regan \&
Teuban 2004) indicate that gas build-up, and hence star formation is
most likely to happen in the central kpc, rather than throughout the
galactic disk.  Similar predictions exist for the gas flows in
galaxy-galaxy interactions and observations of close galaxy pairs
support the prediction of nuclear star formation (Barton et al. 2000;
Ellison et al.  2010; Patton et al. 2011).  We therefore repeat the
above experiment with fibre SFRs, which sample the inner few kpc of
the galaxy\footnote{For $0.02<z<0.1$ the 3 arcsecond fibre diameter
corresponds to a physical size of 1.22 -- 5.53 kpc}.

The SFR-mass relation for unbarred galaxies is re-fit for the fibre
values of both SFR and mass, where the latter are derived from fibre
magnitudes.  In the bottom panel of Figure \ref{delta_sfr} we plot the
$\Delta$ SFRs calculated from the fibre values, although the total
stellar mass is still plotted on the x-axis.  Using the fibre SFRs
allows us to focus on the central part of the galaxy and also
alleviates any residual uncertainty in the aperture corrections.  The figure
shows that for the inner galactic regions sampled by the fibre, there
is an enhancement of around 60\% (0.2 dex) in the SFRs in barred
galaxies compared to the unbarred sample.  However, this enhancement
is only seen in barred galaxies with total stellar masses M $>
10^{10}$ M$_{\odot}$.  The positive $\Delta$ SFR seen in the fibre,
but not the global metric indicates that the enhanced star formation
is both centrally located ($g$-band covering fractions for our
galaxies are typically $\sim$ 15\%) and occuring at a sufficiently
modest rate (on average) that it doesn't drastically affect the
overall galactic SFR.  As noted in Section \ref{sample_sec}, we 
have checked that this is not the result of a redshift bias by
checking that the result remains even with a redshift cut of $z<0.06$
(the range up to which Nair \& Abraham 2010a confirm a constant bar
fraction).  Moreover, the generally high fraction of bars in low
mass galaxies indicates that bars are not systematically missed
below $10^{10}$ M$_{\odot}$ due to other effects such as dust.

\section{Metallicities}\label{mzr_sec}

We apply the same offset technique to metallicities
as a function of mass.  The total stellar mass-metallicity relation of
unbarred galaxies is fit with a second order polynomial and $\Delta$
O/H is defined as the offset between the metallicity of a barred
galaxy and the value predicted for its total stellar mass:

\begin{equation}
\Delta \rm{O/H} = \log \rm{O/H} - \log \rm{O/H}_{\rm predict}
\end{equation}

The O/H residuals (in bins of stellar mass) of the unbarred control sample
have median values around zero and scatter typically less than 0.02 dex.

Note that, unlike the SFRs, no attempt is made to correct the abundance
value to a total metallicity.  The metallicity values used throughout
this paper are fibre values.  Kewley et al. (2005) found that covering
fractions of $>$20\% should yield abundances that are representative
of integrated light spectra over the entire galaxy.  The majority of
our galaxies have fibre covering fractions $<$20\%, so that imposing
such a cut would leave a sample hampered by small number statistics.
However, considering the fibre stellar mass-metallicity relation
circumvents the issue of aperture bias, under the assumption that
the unbarred galaxies have a similar distribution of radial coverage
to the barred sample.  The consistent distributions of mass and
redshift between the barred and unbarred galaxies indicate that this
is a reasonable assumption.

\begin{figure}
\centerline{\rotatebox{0}{\resizebox{8cm}{!}
{\includegraphics{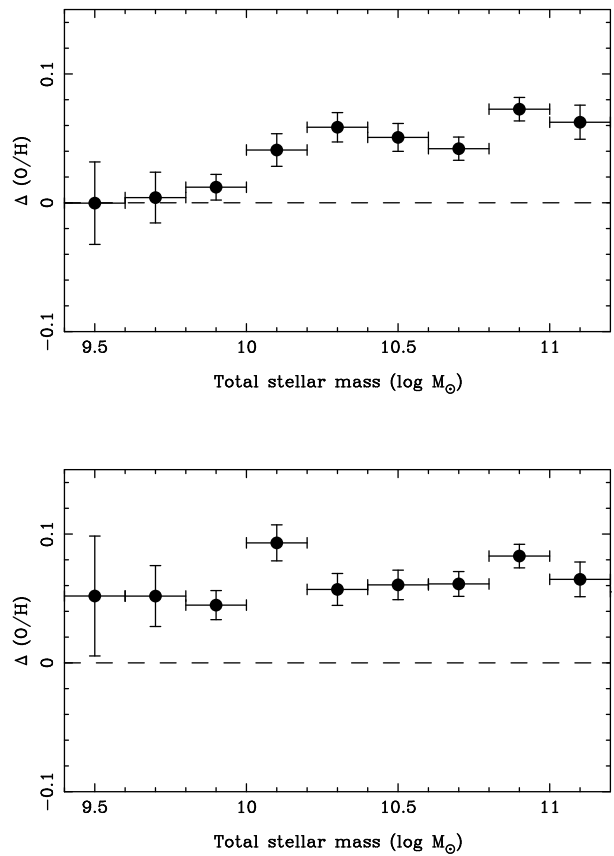}}}}
\caption{\label{delta_oh} The metallicity offset of barred spirals
from the mass-metallicity relation of unbarred spirals.  The top panel
shows the metallicity offset between barred and unbarred galaxies at a
given total galactic stellar mass (as a function of total galactic
stellar mass).  The lower panel shows the metallicity offset between
barred and unbarred galaxies at a given fibre stellar mass (as a
function of total galactic stellar mass).  The increase in $\Delta$
O/H at small stellar masses in the lower panel is due to the smaller
fraction of stellar mass in the SDSS fibre below total stellar masses
of $\sim 10^{10}$ M$_{\sun}$ (Figure \ref{delta_mass}).  }
\end{figure}

\begin{figure}
\centerline{\rotatebox{270}{\resizebox{6cm}{!}
{\includegraphics{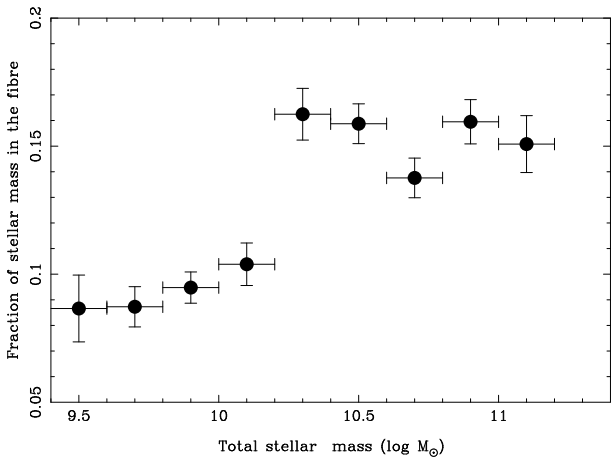}}}}
\caption{\label{delta_mass} The fraction of stellar mass in the SDSS
fibre for barred galaxies as a function of total stellar mass.  Barred
galaxies with M$_{\star}$ $<$ 10$^{10}$ M$_{\odot}$ are dominated by late-type
spirals with relatively small bulges and hence a lower fraction of the
total galaxy mass is in the fibre. Barred galaxies with M$_{\star}$ $>$ 10$^{10}$
M$_{\odot}$ are dominated by early-type spirals with relatively large
bulges (see Figure \ref{bar_images} for examples) so that the fibre
contains a larger fraction of the total galaxy mass.}
\end{figure}

The upper panel of Figure \ref{delta_oh} shows $\Delta \rm{O/H}$ as a
function of total stellar mass.  The metallicities of barred galaxies
are higher than the unbarred sample by $\sim$0.06 dex when the total
mass of the galaxy M$_{\star}$ $>$ 10$^{10}$ M$_{\odot}$.  This is a
similar transition mass to the enhanced star formation rates in Figure
\ref{delta_sfr}.  The lower panel of the same figure shows the offset
from the \textit{fibre} mass-metallicity relation.  The metallicities
of barred galaxies are now higher by $\sim$ 0.06 dex across the entire
mass range.  The appearance of enhanced metallicities at low masses
when fibre quantities are considered can be understood by plotting the
fraction of mass in the fibre as a function of total stellar mass
(Figure \ref{delta_mass}).  Since barred galaxies with M$_{\star}$ $<$
10$^{10}$ M$_{\odot}$ are dominated by late-types (Nair \& Abraham
2010b) whose bulges and bars are small compared to their disks, the
fibre contains a smaller fraction of the total galaxy mass (see also
Figure \ref{bar_images}).  Above M$_{\star}$ $>$ 10$^{10}$
M$_{\odot}$, there is a rapid transition to a population dominated by
early-type spirals, whose bulge fraction is much higher. The low
fraction of mass covered by the fibre for low stellar mass galaxies
results in a more pronounced difference in the fibre mass-metallicity
offsets in the lower panel of Figure \ref{delta_oh}.

\section{Bar length and axial ratio}\label{length_sec}

\begin{figure*}
\centerline{\rotatebox{270}{\resizebox{12cm}{!}
{\includegraphics{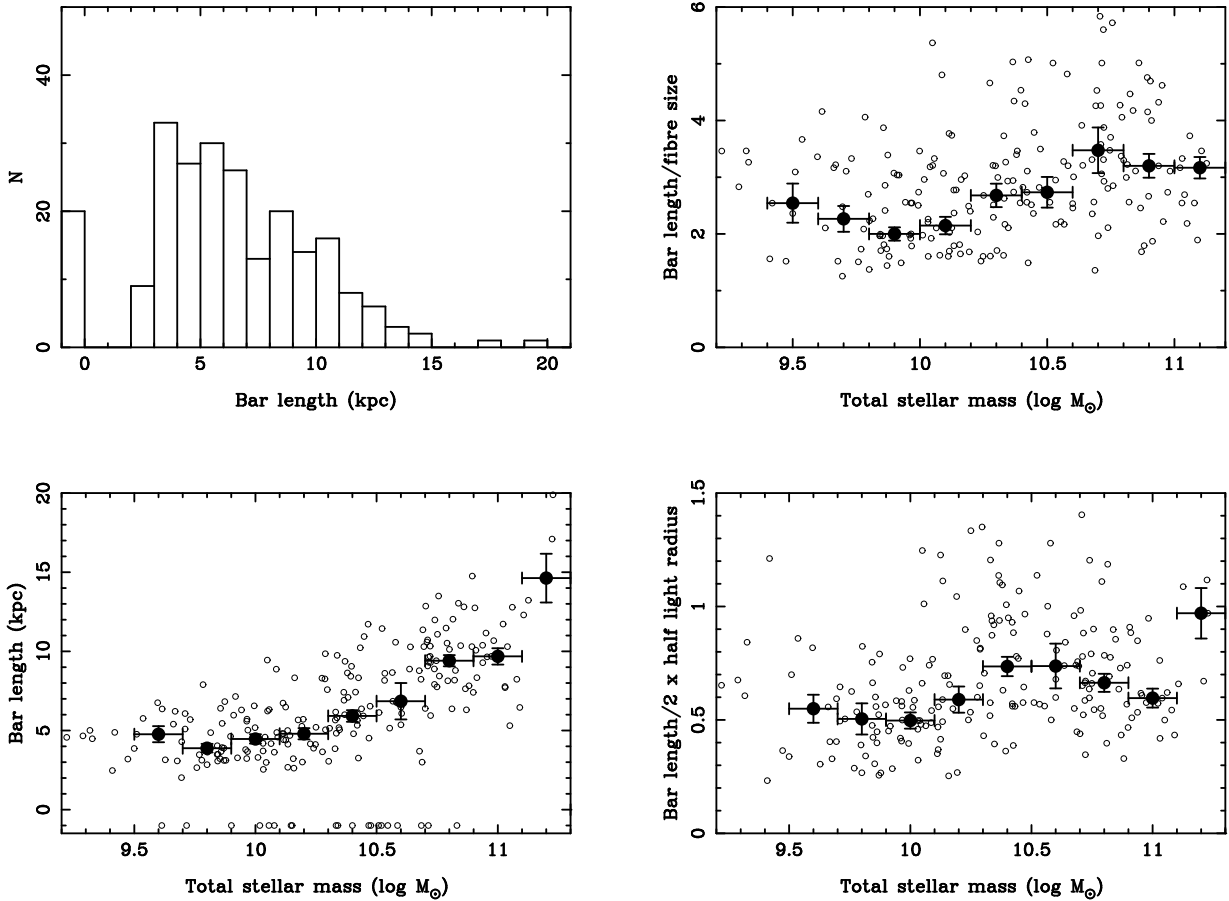}}}}
\caption{\label{lbar} Top left: distribution of bar lengths measured
in the $g$-band.  Values of $-1$ indicate bars for which ellipse
fitting failed and a bar length could not be determined.  Top right:
Ratio of bar length and fibre size.  The bar length is typically 2--4
times larger than the fibre diameter (3 arcseconds).  Bottom left:
Bar length as a function of total stellar mass.  Bottom right:
Fractional bar length as a function of total stellar mass.  In all
panels open points show individiual galaxies and filled points are
medians in stellar mass bins.}
\end{figure*}

The results in Section \ref{sfr_sec} indicate that gas flows in barred
galaxies strongly depend on stellar mass.  Barred galaxies with masses
M $>10^{10}$ M$_{\sun}$ show higher central star formation rates and
higher metallicities.  Galaxies at lower stellar masses, when
corrected for covering fraction, show similar increases in
metallicity, but without the accompanying SFR enhancement.  Clues to
this mass dependence come from the work of Nair \& Abraham (2010b) who
have shown that the fraction of galaxies with bars shows a strong
dependence on stellar mass, which in turn is linked to the
predominance of late type barred galaxies at M $< 10^{10}$ M$_{\sun}$
and early types at higher masses\footnote{The actual fraction of
barred galaxies remains a contentious issue, and is likely due to a
variety of visual and automated classification techniques, and whether
classification is done in the optical or infra-red e.g. Masters et
al. (2010).}.  The changes in metallicity and SFR may therefore also
be linked to position in the Hubble sequence.  Late type barred
galaxies tend to have shorter, weaker bars than early-types (Elmegreen
\& Elmegreen 1985, 1989; Martin 1995; Chapelon et
al. 1999). Simulations indicate that the amount of gas involved in
radial inflows depends on bar length and strength, with the longer,
stronger bars transporting more material at a faster rate than short,
weak bars (Friedli \& Benz 1993; Regan \& Teuban 2004).  Observations
generally support these predictions.  Martin (1995) found that the
majority of starbursting barred galaxies have small axial ratios
(i.e. the bars are strong).  Martinet \& Friedli (1997) also found
that only the long, strong bars in their sample of barred late types
showed signs of enhanced star formation, although they point out that
even strong bars may be in a pre- or post-starburst phase with
`normal' SFRs.  The same is apparently true of early type bars:
Chapelon et al.  (1999) find both starbursting and normal galaxies in
their strongly barred sample.

Complete details on the measurement of bar sizes and axial ratios
will be provided in a future paper (Nair et al, in prep).  In summary,
we use the frequently used procedure based on fitting ellipses to
isophotes (Friedli et al 1996; Jogee et al. 1999; 2004, Abraham et
al. 1999; Abraham \& Merrifield 2000; Sheth et al. 2003, 2008;
Marinova \& Jogee 2007; Menedez-Delmestre et al. 2007). The original
reduced SDSS frames are used to re-extract galaxies with SExtractor
(Bertin \& Arnouts 1996) and the segmentation of each image is checked
to ensure `parent' and `child' objects are correctly separated. For
those galaxies visually classified as barred and with galaxy axial ratios
$>$0.4, we calculate an ellipse from the second order moment of the
light distribution in the $g$-band, stepping from the outer ellipse
towards the centre in sigma bins. For each galaxy we over-plot the
ellipses on the galaxy image and determine the best fitting ellipse by
eye, which normally corresponds to a maximum in the radial profile of
ellipticity and a plateau in the radial profile of the position
angle. The semi-major axis of the fitting ellipse is taken to be the
bar length.

In Figure \ref{lbar} we show the distribution of bar lengths in our
sample. We plot both the absolute value of bar length (in kpc) and the
length relative to the half light radius (taken from Simard et
al. 2011). Cases where the bar length could not be measured (usually
because of its small scale) are allocated lengths of $-1$ in the
Figure.  The non-measurements are included in the calculation of
binned medians shown in the lower two panels of Figure \ref{lbar}.
Most of the bar lengths are between 3 and 12 kpc in the $g$-band, and
the median value is 5.9 kpc.  In all cases, the bars are larger than
the size of the SDSS fibre (upper right panel of Figure \ref{lbar}),
so that the spectra always sample within the co-rotation
radius\footnote{Bars form within the co-rotation radius and
observations of bar pattern speeds indicates that most bars end just
within co-rotation (e.g. Aguerri et al. 2003).}.  We note that there
is little variation in the bar covering fraction in the mass range
$10^{9.5} < $ M$_{\star} < 10^{10.5}$ M$_{\odot}$, indicating that
transition to enhanced fibre SFRs at M $ \sim 10^{10}$ M$_{\odot}$
shown in Figure \ref{delta_sfr} is not due to aperture bias.

In agreement with previous studies, we find that the more massive
(early-type) barred galaxies tend to have longer bars.  This is
true both in an absolute sense (lower left panel of Figure \ref{lbar})
and relative to the half light radius (also measured in the $g$-band)
of the galaxy (lower right panel of Figure \ref{lbar}).  We now
consider whether the SFR enhancement in our barred sample is linked to
the length of the bar.  Previous analyses have tackled this issue by
comparing the structural parameters of starbursting and `normal'
barred galaxies.  Our analysis technique differs from these earlier
studies because we do not simply classify galaxies as starbursting or
normal, but quantify the extent to which star formation is enhanced
($\Delta$ SFR).  This allows us to look for trends between star
formation rates and bar properties, rather than simply comparing the
distributions of two samples.

\begin{figure}
\centerline{\rotatebox{0}{\resizebox{8cm}{!}
{\includegraphics{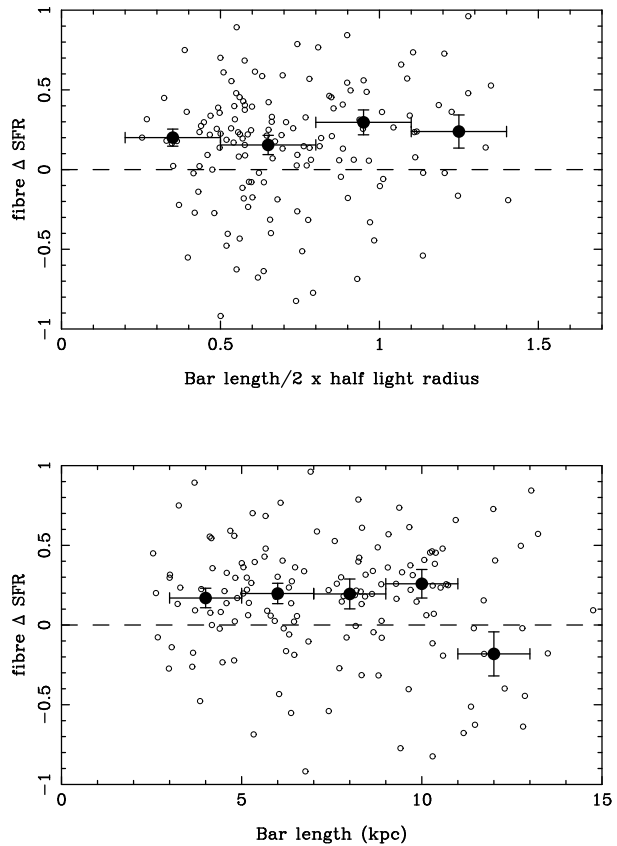}}}}
\caption{\label{delta_lbar} SFR enhancement as a function of absolute
(lower panel) and relative (upper panel) bar length.  Open points
are values for individual galaxies and solid points are binned medians.
There is no correlation between bar length and SFR enhancement. }
\end{figure}

When investigating whether there is a correlation between bar length
and enhanced SFR, we consider only galaxies with M $> 10^{10}$
M$_{\odot}$, which is the regime in which the enhanced rates are
detected.  Figure \ref{delta_lbar} shows $\Delta$ SFR as a function of
relative and absolute bar length.  The lack of correlation indicates
that either bar length is not the most important factor in determining
gas inflow efficiency, or there is some lower limit to the bar length
for inflows that all of our sample exceeds.  Note also the significant
scatter in $\Delta$ SFR at a given bar length, supporting previous
conclusions that the presence of a bar does not gaurantee enhanced
star formation.  There is a broad correlation between bar axial ratio,
b/a, (which is often used as an indicator of bar `strength') and bar
length.  However, the resolution of SDSS images makes it difficult to
measure small axial ratios accurately.  Nuclear structures such as
inner rings can also complicate the fitting procedure.  Visual
inspection of the axial ratio fits yields a sample of 229 acceptable
bar b/a measurements.  Just as there is no dependence of $\Delta$ SFR
on bar length, we also find no correlation with axial ratio.

\section{Discussion}

\subsection{Star formation rates}

It is well established that star formation enhancements in barred
galaxies are mostly seen amongst the early-type population (e.g. Huang
et al. 1996; Ho et al. 1997).  One of our principal results is that
the enhancement in SFR in barred galaxies can be associated with the
more fundamental parameter of galactic stellar mass.  The increase in
SFR in bars is only evident in our sample at M$_{\star}$$>$ 10$^{10}$ M$_{\odot}$.
Earlier conclusions of morphology-dependent SFR enhancements follow
from this mass dependence, since barred galaxies with M$_{\star}$ $>$ 10$^{10}$
M$_{\odot}$ are dominated by early-types. Our results highlight that
the dependence of morphological type on mass can complicate the
interpretation of trends in barred galaxies.  For example, Aguerri
(1999) found that bar strength correlates with star formation rate.
However, as discussed above, bar strength is known to correlate with
Hubble type: low mass late-types having generally weaker bars than
high mass early-types.  The correlation of SFR and bar length found by
Aguerri (1999) may therefore simply be a reflection of the mass-SFR
relation.  Our results emphasize the importance of looking for trends
at a given stellar mass, and we have tackled this by calculating SFR
offsets from the unbarred SFR-mass relation.  In contrast to Aguerri
(1999), we do not find a correlation between SFR enhancement and bar
length.  Instead, there seems to be a mass threshold of M$_{\star}$ $\sim$
10$^{10}$ M$_{\odot}$ above which fibre SFRs are higher, on average, by
$\sim$ 60\% than unbarred galaxies of the same stellar mass.  This
threshold corresponds to the mass at which the barred population
transitions from late to early type barred galaxies.  The nature of
the bar itself changes at this morphological boundary, the late-types
being exponential and the early-types being flat (Elmegreen \&
Elmegreen 1985).  Clues to understanding the differences in bars can
be gleaned from simulations.  Combes \& Elmegreen (1993) found that
late-type bars stop growing soon after their formation, whereas bars
in early-types continue to gain in strength and length.  The morphological
distinction is driven (in the simulations) by the more concentrated
mass profiles of early-types, whose bulges are more prominent than
in the late types.  Galaxies with high bulge fractions have their
co-rotation radius well within the galactic disk and can therefore
effectively transfer angular momentum outwards and grow the bar.  In turn,
a growing bar has continued access to the disk's gas supply, so that 
early-type bars may experience extended inner star formation, 
regardless of the current length of the bar.  Observational support for
these theoretical predictions comes from a correlation between bar
length and bulge fraction (Gadotti 2011).

\subsection{Metallicities}

We have found the ostensibly surprising result that barred galaxies
have higher central metallicities for their (fibre) mass by about 0.06
dex, relative to unbarred galaxies of the same mass.  Previous studies
have concluded that the central gas-phase abundance of barred galaxies
is either lower than, or consistent with unbarred galaxies (Dutil \&
Roy 1999; Henry \& Worthey 1999; Considere et al. 2000).  Before
considering how these results may be reconciled, we point out some
practical improvements that our study offers.  First, we have a sample
of control (unbarred) galaxies which is consistent in mass and
redshift distribution with our barred galaxies.  Selection effects are
therefore largely mitigated.  Our sample spans a relatively large
range in mass and samples both early and late-type barred spirals,
whereas many previous surveys include mostly gas-rich late-types.  Our
sample is large: 294 barred galaxies and 588 unbarred galaxies as a
control.  We search for abundance trends as a function of stellar
mass, whereas previous work has relied mostly on trends with
luminosity.  As shown by Ellison et al. (2008a), when galaxies exhibit
enhanced star formation (as we believe the higher mass barred galaxies
do) the luminosity-metallicity relation can be affected by changes in
both metallicity and luminosity. Finally, our abundances are
determined identically for every galaxy, so that we do not have to be
concerned with either metallicity zero-point offsets (Kewley \&
Ellison 2008) or differences in gradients that may be introduced by
different strong line methods (Moustakas et al. 2010).  All of these
issues mean that our study is likely to offer the most internally
consistent and statistically robust sample for the investigation of
(integrated) metallicities and SFRs in barred galaxies.

It has been recently demonstrated that the mass-metallicity relation
for the general star-forming galaxy population is itself modulated by
SFR, such that galaxies with higher SFRs tend to have lower
metallicities (Ellison et al. 2008b).  Mannucci et al. (2010) and
Lara-Lopez et al. (2010b) have even suggested a fundamental relation
between SFR, mass and metallicity in star-forming galaxies that can be
fit with a plane.  Interestingly, barred galaxies do not
(qualitatively) follow this general trend, since they have both
enhanced SFRs and higher metallicities for their mass, and would
therefore presumably be outliers on the `fundamental relation'.

We now consider what physical reasons may lead to an apparent
discrepancy between our metallicity results (enhancement in barred
galaxies at a given mass) with past studies (low central abundances in
barred galaxies at a given luminosity).  We are guided in this
endeavour by predictions from numerical simulations.  For example,
Friedli et al. (1994) study the evolution of abundance gradients after
bar formation and distinguish between cases with and without star
formation.  They show that in the absence of star formation, radial
mixing leads to a flattened gradient due to both a lower abundance
within the bar co-rotation radius and a higher metallicity at larger
radii due to the effect of outward mixing.  If star formation is
included, the outer region is again relatively metal-rich with a flat
gradient, just as in the no star formation case.  However, inside the
co-rotation radius, the extra star formation leads to chemical
enrichment, steepening the gradient.  In fact, the inner gradient is
largely indistinguishable from the original (pre-bar) metal
distribution, with the exception of strong metal enhancement (by
$\sim$ 0.6 dex in the models of Friedli et al. 1994) in the inner kpc.
All of our fibre metallicities sample regions well within the bar
itself, so for interpreting our results we need only consider the
regime within co-rotation.  Hence, gas flows with no accompanying star
formation should lead to lower metallicities and with star formation
we would expect a metallicity enhancement, assuming there has been
sufficient time for the metals to be returned to the ISM.

Let us now consider the observational results which can be compared to
these theoretical predictions.  Most previous studies have focussed on
late-type (low mass) barred galaxies, in which we find no bar-induced
star formation excess, so the simulations would predict relatively
flat, unbroken gradients.  This is indeed the case, with only a
fraction of barred galaxies exhibiting abundance gradient breaks
(Martin \& Roy 1992; Zaritsky et al. 1994; Martin \& Roy 1995; Walsh
\& Roy 1997; Martin, Lelievre \& Roy 2000).  In these cases,
extrapolating the gradient to zero galactocentric radius would indeed
yield values lower than their unbarred counterparts.  So, why do we
find high abundances in the low mass (late-types) bars in the SDSS,
even in the absence of higher SFRs?  This can not be a covering
fraction effect whereby we sample the metal-enriched gas beyond
co-rotation, because the fibres always sample well within the bar
length (Figure \ref{lbar}).  One possible explanation for the high
metallicities at low masses builds on the scenario described above for
the lack of bar growth in late types (Combes \& Elmegreen 1993).  The
high metallicities, but normal SFRs in the low mass (late-type) bars
in our sample may be due to an early episode of gas inflow
and star formation soon after bar formation.  However, the gas
reservoir available for inflow is quickly depleted, but we now see the
enrichment of that early burst.
 
At masses M$_{\star}$ $>$ 10$^{10}$ M$_{\odot}$, where our sample is
dominated by early type barred spirals, we find enhanced SFRs. The
model of Friedli et al. (1994) therefore predicts that the integrated
metallicity within the bar exceeds the unbarred value.  The models
also predict that in this regime abundance gradient breaks (flatter
beyond co-rotation and steeper towards the galactic centre) should be
commonplace in galaxies with high stellar masses (preferentially
early-types).  The majority of galaxies studied in past abundance work
are late-types where the gas fraction is relatively high and the
number of HII regions is large. Such samples may be dominated by
galaxies without the higher central SFRs that lead to gradient
breaks. Dutil \& Roy (1999) measured relatively flat gradients in 8
early-type barred galaxies, but did not report breaks. However, in
most cases the inner kpc of the galaxies were not well-sampled and the
abundance scatter at a given radius is typically 0.4 dex.  In one
detailed case study (of NGC 4900), Cantin et al. (2010) find bright
emission line regions and gas phase abundances of twice solar only a
few hundred pc from the galactic centre.  Modern instrumentation and
re-calibrated strong line abundance diagnostics have significantly
reduced measurement error and diagnostic uncertainty, yielding tighter
gradients (e.g. Moustakas et al. 2010).  It would be interesting to
re-visit abundance gradients for a large sample of barred galaxies,
focussing on the inner regions.

The high SFRs and high metallicities in barred galaxies with M$_{\star}$ $>$
10$^{10}$ M$_{\odot}$ stand in contrast to close pairs of galaxies which
trace interaction-induced gas flows.  Although enhanced star formation
rates in pairs are ubiquitously reported in the literature, the
metallicities are \textit{lower} at a given mass (Ellison et al. 2008a;
Michel-Dansac et al. 2008).  This may be because galaxies that are
still easily identifiable as pairs (as opposed to post-mergers) are
relatively early on in the dynamical process.  Simulations predict
that the gas-phase metallicity does not increase as a result of the
triggered star formation until hundreds of Myr after the starburst,
by which time the merger may be almost complete (Montuori et al. 2010).  
This is because of the time delay of both
the supernovae and also the cooling time of the gas to reach the
$\sim$ 10$^4$ degrees to which the nebular emission lines respond.
Since the high mass barred galaxies are metal-enhanced, the
process which triggered the star formation (which we still see to be
on-going) must be long-lived ($> 10^8$ years).  This is consistent
with the reconstruction of star formation histories in barred
galaxies that indicate
multiple episodes of central star formation (e.g. Cantin et al. 2010).

\subsection{Comparison with galaxy-galaxy pairs}\label{pairs_sec}

\begin{figure}
\centerline{\rotatebox{270}{\resizebox{6cm}{!}
{\includegraphics{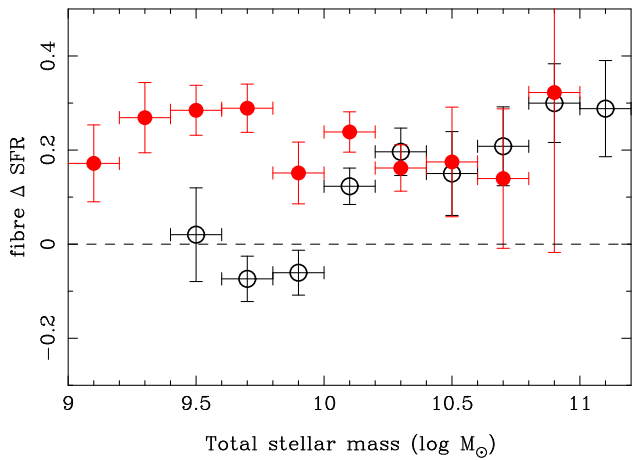}}}}
\caption{\label{sfr_bars_pairs}Enhancement in
fibre SFR as a function of stellar mass.
Open points are for barred galaxies (data points are the
same as for Figure \ref{delta_sfr}) and filled points are for
galaxies with a close companion within 300 \kms\ and a projected
separation of 30 kpc.  }
\end{figure}

Since the two main mechanisms for torquing gas to the centres of
galaxies are usually considered to be the presence of a bar and
galaxy-galaxy interactions (e.g. Mihos \& Hernquist 1996; Di Matteo et
al. 2007), it is of interest to compare the relative impact of these
two processes.  Although bars can themselves be the result of an
interaction, we note that none of the barred galaxies in our sample
appear to be currently undergoing any interaction (or has a close companion 
within 30 kpc).  Bars may therefore
represent a much more extended phase in the galaxy's history than a
fly-by or merger (as supported by the observations cited in this
paper), as well as having a contribution from secular formation
mechanisms.  Our comparison is therefore also one of timescales,
comparing the effect of a close encounter which is short-lived but
potentially dramatic, and the longer-lived bar phase (which may be
either interaction-induced or secular).  We will explicitly address
the formation of bars in close galaxy pairs in a future paper where we
look at the fraction of close pairs in which a bar is detected (Nair
et al. in preparation).

It has not previously been possible to compare the relative
efficiencies of bars versus interactions, which requires a large
sample of homogeneous data for both barred galaxies and galaxies with
close companions.  We can now tackle this comparison for the first
time, by comparing the SDSS study of barred galaxies presented here
with the SFRs that we have previously published for SDSS galaxy pairs
(Ellison et al. 2008a, 2010).  The galaxy pairs are selected from the
sample presented by Patton et al. (2011), compiled from the DR7.  The
stellar masses of the galaxy and its companion are within a factor of
3 and the upper redshift cut is $z=0.1$ (which matches the limit of
the bars sample).  We select the closest pairs in order to yield a
sample most likely to yield enhanced SFRs.  The redshifts must
therefore agree to within a $\Delta V$ $<$ 300 \kms\ and the projected
separation is within 30 kpc.  The control sample is matched in stellar
mass and redshift.

In Figure \ref{sfr_bars_pairs} we show the change in fibre SFR as a
function of mass for both galaxies with a close companion and barred
galaxies (the latter are the same as the points from Figure
\ref{delta_sfr}). The level of enhancement is consistent with the
findings of Ellison et al. (2008a) who found aperture corrected
enhancements of $\sim$ 60\% in their 1:3 mass ratio sample.  The
maximum increase in the fibre star formation rates of both barred galaxies and
galaxies with a close companion is around 0.2--0.3 dex (60--100\%),
indicating that these two processes can be equally efficient.
However, the mass dependence on fibre SFR is very different between
the pairs and the bars.  In Section \ref{sfr_sec} we discussed how the
increase in SFRs in barred galaxies is strongly mass-dependent,
showing a sharp transition at M$_{\star}$ $\sim$ 10$^{10}$ M$_{\odot}$.  This is
not seen in the pairs (where the mass is that of the individual
galaxy), with enhancements up to a factor of two in the fibre SFR down
to total stellar masses M$_{\star}$ $\sim 10^{9}$ M$_{\odot}$.  One
interpretation of Figure \ref{sfr_bars_pairs} is that galaxy-galaxy
interactions and bars can trigger star formation in different mass
regimes.  However, the enhanced metallicity in low mass bars indicates
that their past SFRs were enhanced, but the enhancement was
short-lived.  Therefore, the detection of a SFR enhancement in pairs
of all stellar masses is more likely due to them being observed in the
brief window of triggered star formation which is usually strongest
after the first pericentric passage (e.g. Montuori et al. 2010).

Although bars and galaxies with a close companion appear to have
similar triggering efficiencies, the contribution of these processes
depends on the frequency of bars and pairs.  We can make a simple
estimation of the relative contribution of bars and pairs to the fibre
SFR enhancements ($\epsilon_{b/p}$) in galaxies from

\begin{equation}\label{c_eqn}
\epsilon_{b/p} = \frac{f_{b}}{f_{p}} \times \frac{f_{b,\star}}{f_{p,\star}} \times \frac{10^{\Delta SFR_b}}{10^{\Delta SFR_p}}.
\end{equation}

The first ratio is the fraction of the galaxy population in bars and
pairs ($f_{b}$ and $f_{p}$ respectively).  In order to calculate
$f_{p}$, we use the statistics of Patton \& Atfield (2008), but scaled
to our different pair criteria (mass ratio, projected separation and
$\Delta V$).  This fraction corrects for fibre incompleteness, but not
interloper fraction, since the latter effect is also present in our
estimate of the SFR enhancement.  We determine $f_p$ = 0.024.  Nair \&
Abraham (2010b) quote the fraction of bars in their sample to be
$\sim$ 30\%.  However, this is a fraction of the \textit{disk}
population with bars, whereas we require the fraction of the
\textit{total} population that has a bar.  Imposing a redshift cut of
$z>0.02$ (which is part of the metallicity criterion) and galaxy
b/a$>$0.4 on the Nair \& Abraham (2010a) sample yields 1657 bars out
of a total of 9914 moderately inclined galaxies, leading to $f_{b}$=0.167.  We
note that this is likely to be a conservative fraction as bar
incidences tend to be fairly low in visual classifications of optical
images.  The second component in Eqn \ref{c_eqn} is the ratio of the
fraction of bar and pair galaxies that pass the emission line
criterion (Section \ref{sample_sec}) to enter into our sample of
star-forming galaxies ($f_{b,\star}$ and $f_{p,\star}$ respectively).
The ratio of these fractions accounts for the different selection
functions of bars and pairs, such as differences in covering fraction
and mass distributions which can affect the typical strength of
emission lines.  The number of galaxies in close pairs ($r_p <$ 30 kpc
and $\Delta V < 300$ \kms) with $0.02<z<0.1$ in the sample is 1770, of
which 431 have emission lines strong enough to be included in our
star-forming sample, leading to $f_{p,\star}$ = 0.24.  Of the original
$0.02<z<0.1$ bar sample of 1657, 294 are in our star-forming sample.
However, we make a further conservative decision to only count the
bars that contribute to the \textit{on-going} enhanced star formation,
i.e. those with M $>10^{10}$ M$_{\odot}$, of which there are 212.  We
have argued above that there may have been a past enhancement in the
SFR of lower mass bars, which we can not constrain.  We hence
determine $f_{b,\star}$ = 0.128.  The third ratio in Eqn \ref{c_eqn}
is the ratio of the SFR enhancements in bars and pairs.  Although
there is some mass dependence seen in Figure \ref{sfr_bars_pairs}, on
average this ratio $\frac{10^{\Delta SFR_b}}{10^{\Delta SFR_p}} \sim
1$ (recall that we are only including the bars with M$_{\star}$ $>$
10$^{10}$ M$_{\odot}$ in our calculation). Combined, these adopted
values lead to $\epsilon_{b/p} \sim 3.5$, indicating that
approximately 3.5 times more central star formation is triggered by
the presence of a bar than by interactions.  This value of $\epsilon$
is quite sensitive to the selection criteria, e.g. the definition of a
close pair and how the bar fraction is calculated.  None the less,
$\epsilon_{b/p} \sim 3.5$ is probably a conservative estimate for the reason
described above and indicates the important role of bars in the
build-up of central stellar mass.

We emphasize that our results can be used to calculate the relative
contributions of bars and pairs to
the \textit{central} star formation, due to the low covering fractions
in the bars sample.  Neither bars nor pair interactions appear to
induce additional star formation in the disk (e.g. Ellison et
al. 2010; Patton et al.  2011).

\section{Summary}

We have compiled a sample of 294 galaxies with a large-scale bar and a
control sample of 588 unbarred galaxies visually classified from the
SDSS DR4. All of the galaxies have measurements of stellar mass, star
formation rate and gas phase metallicity.  The sample is used to
study differences in star formation rates and chemical abundances
between the visually classified barred and unbarred populations in
order to investigate evidence of bar-induced radial gas flows.  Our
analysis fits the SFR-mass and mass-metallicity relations to the fibre
values of mass, SFR and metallicity of unbarred galaxies and then
measures deviations from these relations in barred galaxies.  Using
the fibre values compensates for differences in the fraction of mass
in the fibre and reveals trends that are absent in global
measurements.  Our observations reveal that:

\begin{enumerate}

\item The central star formation rates of visually classified barred
galaxies with M$_{\star}$ $>$ 10$^{10}$ M$_{\odot}$ are higher than unbarred
galaxies of the same mass by $\sim$ 60\%.  Lower mass barred galaxies
show no increase in their central star formation rates, above the
values observed in unbarred spirals.

\item In contrast to theoretical expectations and inferences from past
observational studies, we find no significant correlation between star
formation rate enhancement and bar length or bar axial ratio in the
galaxies where increased star formation is found (i.e. at M$_{\star}$ $>$
10$^{10}$ M$_{\odot}$).

\item A 60\% enhancement is also typical of the star formation
triggered in galaxy-galaxy interactions at masses M$_{\star}$ $\sim$ 10$^{10}$
M$_{\odot}$.  However, interactions also show SFR enhancements at masses M
$<$ 10$^{10}$ M$_{\odot}$, whereas visually classified barred galaxies
do not.

\item Although the enhancement in the SFRs of visually classified
barred and pair galaxies is similar, the higher fraction of barred
galaxies than pairs means that bars dominate over interactions in
their contribution to the central star formation rates for galaxies
with M$_{\star}$$>$10$^{10}$ M$_{\odot}$.  We estimate that $\sim$ 3.5 times
more central star formation comes from the presence of a bar than from
interactions.

\item In contrast with previous observations that have found low
central gas phase abundances in visually classified barred galaxies,
we find higher metallicities by $\sim$ 0.06 dex relative to unbarred
galaxies at the same mass.  The high metallicities in barred galaxies
also contrast with galaxy pairs where gas inflows lead to low
metallicities.

\item The metal-enrichment is seen even in low mass barred galaxies
where no enhancement in the SFR is observed.  

\end{enumerate}

Taken together, these results suggest a picture of
mass dependent gas inflows in galaxies with strong stellar bars.
The enhanced chemical abundances in barred galaxies of all masses
indicates that most of them have undergone some past enhancement
in their star formation rates.  Due to the time delay for enrichment
following the star formation, this in turn indicates that the
bars themselves are relatively long-lived.  However,
our results are consistent with a distinction between the triggered
star formation histories experienced by low mass, late-type, barred galaxies
and the higher mass, early-types.  Below a mass threshold of M$_{\star}$ $\sim$ 
10$^{10}$  M$_{\odot}$, the star formation that has led to higher chemical
abundances has now ceased, indicating that any SFR enhancement was 
relatively short-lived.  For barred galaxies with stellar masses
M$_{\star}$ $>$ 10$^{10}$  M$_{\odot}$, the enhanced star formation is on-going.

\section*{Acknowledgments} 

Thanks to Francoise Combes, Ken Freeman, Isabel Perez and Patricia 
Sanchez-Blazquez for comments on an earlier draft of this paper.
We are grateful to the MPA/JHU group for
access to their data products and catalogues (maintained by Jarle
Brinchmann at http://www.mpa-garching.mpg.de/SDSS/). 
SLE and DRP acknowledge the receipt of NSERC Discovery grants which
funded this research.  We thank the anonymous referee for an insightful
report.

\end{document}